\documentclass[11pt]{article}
\usepackage{amssymb,amsmath}
\usepackage{float}
\usepackage{color,fullpage}
\usepackage{amsthm}
\usepackage{graphicx}   
\usepackage{algorithm}
\usepackage{algorithmic} 
\usepackage{graphics} 
\usepackage{caption}
\usepackage{subcaption}
\usepackage{multirow}
\usepackage{url}
\usepackage{mathdots}
\usepackage{arydshln}

\newtheorem{theorem}{Theorem}

\newtheorem{lemma}{Lemma}

\newtheorem{definition}{Definition}

\numberwithin{equation}{section}
\numberwithin{figure}{section}
\numberwithin{definition}{section}

\begin{document}
\bibliographystyle{unsrt}
\title{Phase Retrieval with background information}

\author{Ziyang Yuan\thanks{
College of Science, National University of Defense Technology,
Changsha, Hunan, 410073, P.R.China. Corresponding author. Email: \texttt{yuanziyang11@nudt.edu.cn}}
\and Hongxia Wang{\thanks{
		College of Science, National University of Defense Technology,
		Changsha, Hunan, 410073, P.R.China. Email: \texttt{wanghongxia@nudt.edu.cn}}
}
}



\date{}

\maketitle

\begin{abstract}
	Phase retrieval problem has been studied in various applications. It is an inverse problem without the standard uniqueness guarantee. To make complete theoretical analyses and devise efficient algorithms to recover the signal is sophisticated. In this paper, we come up with a model called \textit{phase retrieval with background information} which recovers the signal with the known background information from the intensity of their combinational Fourier transform spectrum. We prove that the uniqueness of phase retrieval can be guaranteed even considering those trivial solutions when the background information is sufficient. Under this condition, we construct a loss function and utilize the projected gradient descent method to search for the ground truth. We prove that the stationary point is the global optimum with probability 1. Numerical simulations demonstrate the projected gradient descent method performs well both for 1-D and 2-D signals. Furthermore, this method is quite robust to the Gaussian noise and the bias of the background information.\\
	$\mathbf{keywords}$:~~phase retrieval,~~uniqueness,~~projected gradient descent,~~background information
	
\end{abstract}

\section{Introduction}
Phase retrieval is to recover phase from the signal's intensity only measurement in some transform domain. In this paper, Fourier transform domain is only concerned. Phase retrieval arises in a variety of applications such as X-ray crystallography, astronomy, coherent diffraction imaging, Fourier ptychography\cite{Miao1999Extending}\cite{fienup1987phase}\cite{Zheng2013Wide}. It can be formulated as:
\begin{eqnarray}
	&\mathbf{Find}~\mathbf{x}&\nonumber\\
	&\mathrm{s.t.}~b_i=|\mathbf{F}_i^*\mathbf{x}|^2,~i=1,\cdots,m,&
\end{eqnarray}
where $\mathbf{b}$ is the Fourier power spectrum,  $\mathbf{x}\in\mathbb{C}^n$ is the signal of interest, $\mathbf{F}_i=(e^{\frac{2\pi ji0}{m}},e^{\frac{2\pi ji1}{m}},\cdots,e^{\frac{2\pi ji(n-1)}{m}})^{\text{T}}$ is the Fourier vector where $j=\sqrt{-1}$ and $(\cdot)^{\text{T}}$ is the transpose.\\
\indent
Without any additional constraints, the solution of (1) isn't unique. Specifically, if $\mathbf{x}=\mathbf{y}e^{j\mathbf{\theta}},~\theta\in[0,2\pi]$, then $|\mathbf{F}_i^*\mathbf{x}|=|\mathbf{F}_i^*\mathbf{y}|,~i=1,\cdots,m$. Except for the global phase, when $m=n$, the inverse conjugate and time translation of $\mathbf{x}$ can also satisfy the constraints in (1). We denote those solutions above as trivial solutions. When mentioning the uniqueness, it often excludes those trivialities. Moreover, the feasible set satisfying (1) isn't convex. These ill conditions above make phase retrieval problem difficult. As a result, many works were come up to deal with the phase retrieval problem.\\
\indent
From the theoretical side, there are lots of works concerning the uniqueness of phase retrieval. When it comes to the Fourier transform, \cite{Hofstetter1964Construction} proves that for the 1-D signal there is no uniqueness guarantee for phase retrieval despite considering these trivialities above. \cite{Hayes1982The} points out that when the dimension of the signal is larger than or equal to 2, the solution of the phase retrieval problem can be unique except for a zero measurement set. Based on those theoretical works, various methods were come up to enforce the uniqueness of phase retrieval so that efficient algorithms can be devised to search for the ground truth.\\  
\indent
The signal of interest $\mathbf{x}$ is often considered to be real and nonnegative with a known support. Under these conditions, error reduction method and Hybrid Input and Output method(HIO)\cite{Gerchberg1971A}\cite{Fienup1982Phase} were come up. Usually, these conditions can't enforce the uniqueness of phase retrieval problem\cite{Beinert2015Ambiguities} besides those methods are liable to be stagnated during the iterations without theories guaranteed to converge to the real solution. But those methods can find an acceptable result and were used widely in practice.\\
\indent
The sparsity of the signal is often utilized as a priority to deal with the phase retrieval problem. Under some proper conditions, the uniqueness of the phase retrieval problem can be guaranteed. In \cite{6853917}, if $m$ is a prime, then $m\geq k^2-k+1$ measurements are sufficient to guarantee the uniqueness of phase retrieval problem. In \cite{JURi2013Unique}, if the auto-correlation sequence of $\mathbf{x}$ is determined besides it is collision free, then $\mathbf{x}$ is sufficient to be uniquely guaranteed if $k\neq6$. In \cite{Shechtman2013GESPAR}, GESPAR method was come up to deal with the sparse phase retrieval problem with a good performance.\\
\indent
In\cite{Eldar2014Sparse}\cite{Bendory2017Phase}, the frames of STFT were also considered. Utilizing the redundant property of STFT frames, the solution of (1) can be unique with proper constraints about the quantities or the bandwidth of the frames. Except for the multiple measurements, the information of the inference signal can also be used to guarantee the uniqueness of phase retrieval in\cite{Kim1990Iterative}\cite{Kim2002Phase}\cite{Raz2013Vectorial}. Interested readers can refer to \cite{Beinert2017Enforcing} which has a comprehensive review about the method to guarantee the uniqueness solution of phase retrieval.\\
\indent
In this paper, we come up with a new model called phase retrieval with the background information. Under this model, projected gradient descent method can be devised to find the solution of phase retrieval. The mathematical formulation of this model is described by (2). The main result of this paper is that the exact uniqueness(even neglecting the trivialities) of the phase retrieval can be guaranteed if one of two situations is satisfied:
\begin{itemize}
	\item  $m\geq2(n+k)-2$, besides $k\geq n$, $y_k\neq0$
	\item  $m=n+k+p$, $p\geq 0$, besides
	\begin{eqnarray*}
		k\geq
		\left\{
		\begin{aligned}
			&\text{max}~(3n-2-p,n),~m-2n+1\text{~is odd}&\\
			&\text{max}~(3n-1-p,n),~m-2n+1\text{~is even}&
		\end{aligned}
		\right..
	\end{eqnarray*}
	and $y_{l}\overset{i.i.d}{\sim}\mathcal{N}(\mu,\sigma^2)$, $l=1,\cdots,k$.
\end{itemize}
\indent
Here $k$ is the length of the background information, $p$ is the oversample size, $y_{l},~l=1,\cdots,k$ are the elements of the background information. When $p=0$, under the second condition we build a loss function and utilize the projected gradient descent method to search for the global optimum besides proving that the stationary point is the global optimum with probability 1. Simulations demonstrate the effectiveness of the algorithm under this model.\\
\indent
This paper is organized as below. In section 2, the uniqueness theories about the phase retrieval with background information are established. In section 3, algorithm to deal with phase retrieval problem with background information is introduced. In section 4, simulation tests are applied to demonstrate the effectiveness of the algorithm. Section 5 is the conclusion.
\section{Phase retrieval with background information}
\indent
In practice, we can put the sample onto a plate which can be a ground glass or something semi-transparent besides the image of the plate is known in advance which is called the background information(see Figure 1). When we get the Fourier modules of this combination, the problem is to recover the object $\mathbf{x}$ from the combinational Fourier power spectrum $\mathbf{b}$ and the background information $\mathbf{y}$.\\
\begin{figure}
	\centering
	\includegraphics[width=4in]{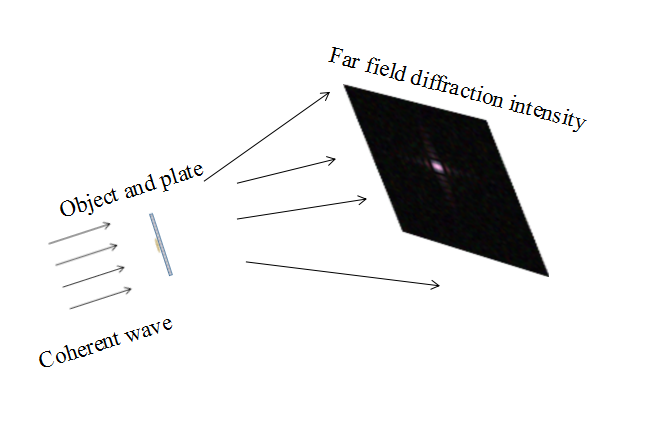}
	\caption{The procedure of phase retrieval with background information.}
\end{figure}
\indent
The mathematical formulation of Figure 1 can be represented as:
\begin{subequations}
	\begin{eqnarray}
		&\mathbf{Find}~\mathbf{z}&\nonumber\\
		&\mathrm{s.t.}~b_i=|\mathbf{F}_i^*\mathbf{z}|^2,~i=1,\cdots,m,&\\
		&z_{n+l}=y_{l},~l=1,\cdots,k.
	\end{eqnarray}
\end{subequations}
Let $\mathbf{z}=(\mathbf{x};\mathbf{y})\in\mathbb{R}^{n+k}$. In this paper, we always assume the solution set of (2) isn't empty. For simplicity, the signal is real in our paper but the theory and algorithm can be extended to the complex signal. We have already known that the solution which satisfies the constraints (2a) is not unique. But we prove that the solution of (2) is unique with proper conditions on (2b) which means (2b) might vastly reduce the feasible set of $\mathbf{z}$ given by (2a). One of the main theorems in this paper is presented as below:
\begin{theorem} 
	When $m\geq2(n+k)-2, k\geq n$, and $y_k\neq0$, (2) has an unique solution $\mathbf{z}$. 
\end{theorem}
\indent
Theorem 2.1 suggests that if the background information is sufficient besides the values of the brim of these background aren't zero, there is only one solution satisfying the constraints in (2). Before proving, we will introduce the auto-correlation of a vector $\mathbf{z}$.\\
\indent
The autocorrelation $\mathbf{a}=[a_{-n+1},\cdots,a_{-2},a_{-1},a_0,a_1,\cdots,a_{n-1}]$ of a vector $\mathbf{z}\in\mathbb{R}^n$ can be calculated as:
\begin{eqnarray}
	a_i=
	\left\{
	\begin{aligned}
		&\sum_{p=1}^{n-1}z_p z_{i+p},~i=0,1,\cdots,n-1&\\
		&\sum_{p=-1}^{-n+1}z_p z_{i+p},~i=-1,-2,\cdots,-n+1&
	\end{aligned}
	\right.,
\end{eqnarray}
where $z_{-p}=z_p$, $p=1,\cdots,n$. $\mathbf{a}$ is symmetric namely $a_i=a_{-i}$.  In order to express (3) simply, we define the auto-correlation matrix.
\begin{definition}
	The auto-correlation of a vector $\mathbf{z}\in\mathbb{R}^{n+k}$ can be expressed as the matrix-vector production $\mathbf{A}_0\hat{\mathbf{z}}$ where $\mathbf{A}_0\in \mathbb{R}^{{(2(n+k)-1)}\times{(2(n+k)-1})}$ is the auto-correlation matrix:
	\begin{eqnarray}
		\centering
		\mathbf{A}_0=\left[
		\begin{array}{ccccccccc}
			0&0&\cdots&z_1&z_2&z_3&\cdots&z_{n+k}\\
			\vdots&\vdots&\ddots&\vdots&\vdots&\vdots&\ddots&\vdots\\
			0&z_1&\cdots&z_{n+k-1}&z_{n+k}&0&\cdots&0\\
			z_1&z_2&\cdots&z_{n+k}&0&0&\cdots&0\\
			z_2&z_3&\cdots&0&0&0&\cdots&z_1\\
			\vdots&\vdots&\ddots&\vdots&\vdots&\vdots&\ddots&\vdots\\
			z_{n+k}&0&\cdots&0&z_1&z_2&\cdots&z_{n+k-1}\\
		\end{array}
		\right],
	\end{eqnarray}
\end{definition}
where $\hat{\mathbf{z}}=[\mathbf{z};\overbrace{\left.0;0;\cdots;0\right.}^{n+k-1}]$. $\mathbf{A}_0$ is a Hankel matrix. So if $z_{n+i}=y_i,i=1,\cdots,k$, by calculating $\mathbf{A}_0\hat{\mathbf{z}}$ and extracting the last $n+k$ items, we obtain:
\begin{eqnarray}
	\left\{
	\begin{array}{ccccccccccccc}
		z_1^2&+&z_2^2&+\cdots+&z_n^2&+&y_1^2&+&y_2^2&+\cdots+&y_k^2&=&a_0\\
		&&z_1z_2&+\cdots+&z_{n-1}z_n&+&z_ny_1&+&y_1y_2&+\cdots+&y_{k-1}y_{k}&=&a_1\\
		&&&&&&&&&&&\vdots&\\
		&&&&&&z_1y_1&+&z_2y_2&+\cdots+&z_ky_k&=&a_n\\
		&&&&&&&&&&&\vdots&\\
		&&&&&&&&&&z_1y_k&=&a_{n+k-1}\\
	\end{array}
	\right..
\end{eqnarray}
\indent
In (5), the first $n$ equalities containing the cross- multiplication between the items $z_i,i=1,\cdots,n$ which can usually lead to the non-uniqueness of (2). Only the last $k$ equalities contains the linear combinations of $z_i,~i=1,\cdots,n$. As a result, the theorem given below ensures the unique solution of (2).
\begin{theorem}
	Assuming the auto-correlation $\mathbf{a}$ of $\mathbf{z}\in\mathbb{R} ^{n+k}$ as shown in (3) is known. If $y_k\neq0$, besides $k\geq n$, then the solution of (2) is unique. 
\end{theorem}   
Proof: If $\mathbf{a}$ is known, extracting the last $k$ equalities from (5) and express it as the type of matrix-vector production as below:
\begin{eqnarray}
	\left(
	\begin{array}{cccc}
		y_k&y_{k-1}&\cdots&y_{1}\\
		0&y_k&\cdots&y_{2}\\
		\vdots&\vdots&\ddots&\vdots\\
		0&0&\cdots&y_k\\
		
	\end{array}
	\right)
	\left(
	\begin{array}{c}
		z_k\\
		z_{k-1}\\
		\vdots\\
		z_{1}\\
		
	\end{array}
	\right)
	=
	\left(
	\begin{array}{c}
		a_n\\
		a_{n+1}\\
		\vdots\\
		a_{n+k-1}\\
		
	\end{array}
	\right).
\end{eqnarray} 
Because the solution set of (2) isn't empty. If $y_k\neq0$, the coefficient matrix of (6) is nonsingular, besides we know that the real solution $z_i$, $i=1,\cdots,k$ is uniquely determined.\\
\indent
As a result, if $k\geqslant n$, $\mathbf{x}$ is thus uniquely determined by (6).\hfill$\blacksquare$\\
\indent
Note that, when $k<n$, we can't ensure the uniqueness of (2). Especially let $k=n-1$, we have:
\begin{eqnarray}
	z_1z_n+z_2y_1+\cdots+z_{n-1}y_{k-1}+z_{n}y_{k}=a_{n-1}.
\end{eqnarray}
Although $z_1,z_2,\cdots,z_{n-1}$ can be derived from (6), $z_n$ can't be determined uniquely if $z_1+y_k=0$.\\
\indent
Now, the problem is how to get $\mathbf{a}$. Based on the Winer-Khinchin Theorem $\mathbf{a}$ is calculated from the inverse Fourier transform of $\mathbf{b}$. When the length of the signal is limited, $\mathbf{a}$ can be obtained by oversampling the spectrum.
\begin{lemma}
	Assuming $n$ and $k$ are given, when the sample size $m\geq2(n+k)-2$, $\mathbf{a}$ can be calculated from the inverse Fourier transform of $\mathbf{b}$.
\end{lemma}
Proof: Oversampling namely $m>(n+k)$, we denote $|\tilde{\mathbf{F}}_i ^*\hat{\mathbf{z}}|^2=b_i$, $i=1,\cdots,m$, where $\tilde{\mathbf{F}}_i=(e^{\frac{2\pi ji0}{m}},e^{\frac{2\pi ji1}{m}},\cdots,e^{\frac{2\pi ji(m-1)}{m}})^{\text{T}}$ is the standard Fourier vector. $\hat{\mathbf{z}}=(\mathbf{z};\overbrace{\left.0;0;\cdots;0\right.}^{m-n-k})$. Construct the circular matrix $\mathbf{A}_0\in\mathbb{R}^{m\times m}$ as the same type of (4), but the first row of $\mathbf{A}_0$ is padded with $m-n-k$ zeros. By the circular autocorrelation of discrete Fourier transform, $\mathbf{a}_0=\mathbf{A}_0\hat{\mathbf{z}}$ can be obtained from the inverse Fourier transform of $\mathbf{b}$.\\ 
\indent
By the definition of $\mathbf{a}$ in (3) and calculating $\mathbf{a}_0$, $\mathbf{a}$ can be obtained from $\mathbf{a}_0$ when $m\geq 2(n+k)-2$. \hfill$\blacksquare$\\
\indent
As a result, Theorem 2.1 is concluded by Theorem 2.2 and Lemma 2.1. Recall the error reduction method in \cite{Fienup1982Phase} where the non-support areas of the signal are all at the brim of the signal, i.e., the values of the background information are all zero. Its auto-correlation may contain many cross items as described in (5). As a result, it can't usually ensure the uniqueness of the phase retrieval. Although similar conclusions were given in \cite{Beinert2015} and \cite{Nawab1983Signal}, we prove Theorem 2.1 here with more details and introduce some knowledge prepared for the latter proof.\\
\indent
Theorem 2.1 demands $m\geq2(n+k)-2$. What if $m<2(n+k)-2$? Under this condition, the auto-correlation $\mathbf
{a}$ generally can't be obtained from $\mathbf{b}$. But the correlation between the signal and the background information can be utilized to ensure the uniqueness.
\begin{theorem}
	When $m=n+k+p$, if $k$ satisfies,
	\begin{eqnarray*}
		k\geq
		\left\{
		\begin{aligned}
			&\mathrm{max}~(3n-2-p,n),~\text{if}~m-2n+1\text{~is odd}&\\
			&\mathrm{max}~(3n-1-p,n),~\text{if}~m-2n+1\text{~is even}&
		\end{aligned}
		\right.
	\end{eqnarray*}
	and $y_l\overset{i.i.d}{\sim}\mathcal{N}(\mu,\sigma^2)$, $l=1,\cdots,k$, solution of (2) is unique with probability 1.
\end{theorem}
\indent 
As the proof in Lemma 2.1 discussed, what we get from the inverse Fourier transform of $\mathbf{b}$ is $\mathbf{A}_0\hat{\mathbf{z}}$. The main idea is to find at least $n$ equalities of the linear combinations of $z_i,i=1,\cdots,n$ to guarantee the uniqueness of (2). The number of these linear combinations indeed relates to the size of the background $k$.
\begin{lemma}
	Given $n$ and $m\geq n+k$, there will be at most $\mathrm{max}(0,m-2n+1)$ linear combinations of $z_i$, $i=1,\cdots,n$.  
\end{lemma} 
Proof: Because the matrix $\mathbf{A}_0$ is a circular Hankel matrix. Observing the vectors (8) and (9). 
\begin{eqnarray}
	&\left(
	z_1,z_2,\cdots,z_n,\right. y_1,y_2,\cdots,y_k,\underbrace{\left.0,0,\cdots,0
		\right)}_{m-n-k},&\\
	&(y_1,y_2,\cdots,y_k,0,0,\cdots,0,z_1,z_2,\cdots,z_n).&
\end{eqnarray}
\indent
The vector (8) making inner product with every left shift of (9) is actually the procedure of $\mathbf{A}_0\hat{\mathbf{z}}$.Supposing $m$ is large enough, to avoid the cross multiplications of $z_i$, $i=1,\cdots,n$, there will be $m-2n+1$ kinds of situations. But depending on the size of the padding zero $p$ and the size of the background information $k$, the coefficients of $z_i,~i=1,\cdots,n$ may be all zero. Thus, there are at most $m-2n+1$ linear combinations of $z_i,~i=1,\cdots,n$. When $m=n+k$, if the size of the background information $k$ is less than $n$, there will be no linear combinations of $z_i,~i=1,\cdots,n$. \hfill$\blacksquare$\\
\indent
If we get enough linear combinations of $z_i$, $i=1,\cdots,n$, besides the coefficient matrix is non-singular, (2) has a unique solution $\mathbf{z}$. Now the sufficient condition is come up to ensure the solution of (2) is unique. Before proving, some notations are given.\\
\indent
When the vector (8) makes inner product with every left shift of (9) which also creates constant terms $\mathbf{c}\in\mathbb{R}^{m-2n+1}$. Accordingly, $\hat{\mathbf{a}}\in\mathbb{R}^{m-2n+1}$ denotes as the sub-vector of autocorrelation $\mathbf{a}$ determined by the linear combinations of $z_i,~i=1,...,n$ and $\mathbf{c}$.\\
\indent
Denote
\begin{eqnarray*}
	\mathbf{H}=\left(
	\begin{array}{cccccc}
		y_1&y_2&\cdots&y_n\\
		y_2&y_3&\cdots&y_{n+1}\\
		\vdots&\vdots&&\vdots\\
		y_{k-n+1}&y_{k-n+2}&\cdots&y_{k}\\
		y_{k-n+2}&y_{k-n+3}&\cdots&0\\
		\vdots&\vdots&\ddots&\vdots\\
		y_k&0&\cdots&0\\
		\\
		\hdashline[2pt/2pt]
		\\
		&\multirow{2}{*}{\huge{$\mathbf{0}$}}&&\\
		\\
	\end{array}
	\right),
	\mathbf{T}=\left(
	\begin{array}{cccccc}
		&\multirow{2}{*}{\huge{$\mathbf{0}$}}&&\\
		\\
		\\
		\hdashline[2pt/2pt]
		\\
		y_k&0&\cdots&0\\
		y_{k-1}&y_{k}&\cdots&0\\
		\vdots&\vdots&\ddots&\vdots\\
		y_{k-n+1}&y_{k-n+2}&\cdots&y_{k}\\
		y_{k-n}&y_{k-n+1}&\cdots&y_{k-1}\\
		\vdots&\vdots&&\vdots\\
		y_1&y_2&\cdots&y_n\\
		
	\end{array}
	\right).
\end{eqnarray*}
where $\mathbf{H}\in\mathbb{R}^{{m-2n+1}\times n}$ is a Hankel matrix, $\mathbf{T}\in\mathbb{R}^{m-2n+1\times n}$ is a Toeplitz matrix. We can obtain the equalities below:
\begin{eqnarray}
	\mathbf{H}\mathbf{z}+\mathbf{T}\mathbf{z}=\hat{\mathbf{a}}-\mathbf{c},
\end{eqnarray}
where $\mathbf{z}=[z_1,z_2,\cdots,z_n]^\mathrm{T}$.
Notice that $\mathbf{H}+\mathbf{T}$ is symmetric along the central line namely $h_{i,j}+t_{i,j}=h_{m-2n+2-i,j}+t_{m-2n+2-i,j}$. Thus, 
\begin{eqnarray}
	\text{Rank}(\textbf{H+T})
	\leq
	\left\{
	\begin{aligned}
		&\text{min}((m-2n+2)/2,n),~\text{if}~m-2n+1\text{~is odd}&\\
		&\text{min}((m-2n+1)/2,n),~\text{if}~m-2n+1\text{~is even}&\\
	\end{aligned}
	\right..
\end{eqnarray}
The solution of (10) is unique if and only if $\text{\text{Rank}}(\mathbf{H}+\mathbf{T})=n$.\\
\indent
Assume $m-2n+1$ is odd, Let $\mathbf{H}_1$ denote the first $(m-2n+2)/2$ rows of $\mathbf{H}$, and $\mathbf{H}_2$ denotes the last $(m-2n+2)/2$ rows of $\mathbf{H}$. Because of the central symmetry of $\mathbf{H}+\mathbf{T}$ besides $\text{\text{Rank}}(\mathbf{H}+\mathbf{T})=\text{\text{Rank}}(\mathbf{H}_1+\hat{\mathbf{I}}\mathbf{H}_2)$, (10) is equivalent to 
\begin{eqnarray}
	(\mathbf{H}_1+\hat{\mathbf{I}}\mathbf{H}_2)\mathbf{z}=\hat{\mathbf{a}}_1-\mathbf{c}_1
\end{eqnarray} 
where $\hat{\mathbf{a}}_1$ and $\mathbf{c}_1$ are the sub-vector of $\hat{\mathbf{a}}$ and $\mathbf{c}$ constituted by the first $(m-2n+2)/2$ elements.
The nonzero elements of $\hat{\mathbf{I}}$ are $\hat{I}_{i,(m-2n+2)/2+1-i}=1,~i=1,\cdots,(m-2n+2)/2$. \\
\indent
(10) has a uniqueness guarantee if and only if (12) has a unique solution. Here, we give the theorem below
\begin{theorem}
	Assume $m-2n+1$ is odd. If $k\geq\text{\textrm{max}}~(3n-2-p,n)$, $p=m-n-k\geq0$, $\mathbf{H}$ in (11) is a random Hankel matrix, e.t., its elements $h_{i,j}\overset{i.i.d}{\sim}\mathcal{N}(\mu,\delta^2)$ in the first row and the last column. The solution of (12) is unique with probability 1. 
\end{theorem}
Proof. The main idea is to prove $\text{\text{Rank}}(\mathbf{H}_1+\hat{\mathbf{I}}\mathbf{H}_2)=n$. As (11) showed, if $\text{\text{Rank}}(\mathbf{H}_1+\hat{\mathbf{I}}\mathbf{H}_2)=n$, then $k\geq3n-2-p$.\\
\indent
When $0\leq p\leq n+k-3$, $\mathbf{H}_1$ is a random Hankel matrix. $\hat{\mathbf{I}}\mathbf{H}_2$ is a Toeplitz matrix. There are $k$ Gaussian random variables in $\mathbf{H}_1+\hat{\mathbf{I}}\mathbf{H}_2$. The Lemma 4.1 in \cite{Pan2012Condition} implies that a nonzero polynomial vanishes with probability 0 if the variables are Gaussian ranging over the real line. There are several instances for $\mathbf{H}_1+\hat{\mathbf{I}}\mathbf{H}_2$ with full column rank $n$. As a result, when  $\mathbf{H}_1+\hat{\mathbf{I}}\mathbf{H}_2$ is a square matrix, it is nonsingular with probability 1 because its determinant is a polynomial of the Gaussian variables. Likewise when $\mathbf{H}_1+\hat{\mathbf{I}}\mathbf{H}_2$ is a rectangular matrix,  it is of full column rank with probability 1. As a result, the solution of (12) has uniqueness guaranteed. To conclude, if $k\geq \text{max}(3n-2-p,n)$, (12) has a unique solution with probability 1.\\
\indent Similar conclusion is also obtained by utilizing the same idea when $m-2n+1$ is even. Then Theorem 2.3 is concluded. Especially when $m\geq2(n+k)-2$ namely $p\geq n+k-2$, (12) includes (6). Because $y_k=0$ with probability 0, utilizing the same idea in Theorem 2.2, if $k\geq n$, the unique solution of (12) is also guaranteed. Under this condition, Theorem 2.3 is actually a special case of Theorem 2.1 when $m\geq2(n+k)-2$.\hfill$\blacksquare$\\
\indent
By the Parseval's Theorem, the solution of (1) is laid on the hyper-ball with radius $\sqrt{\sum_{i=1}^{m}b_i/m}$. Especially, the solution set of (1) is a non-convex set. With proper conditions described above, the set satisfying the constraints in Theorem 2.1 or Theorem 2.4 intersects the hyper-ball with the only solution $\mathbf{x}$ with probability 1.\\
\indent
\textit{Remark}: We notice that some published works about the phase retrieval with inference signal which are closely related to our model in this paper.  in\cite{Kim1990Iterative} it knows the information about the spectrum of the original signal $|\hat{\mathbf{x}}|^2$, inference signal $\mathbf{y}$ and $|\hat{\mathbf{x}}+\hat{\mathbf{y}}|^2$. In\cite{Kim2002Phase}\cite{Raz2013Vectorial}, $|\hat{\mathbf{x}}|^2$, $|\hat{\mathbf{x}}+\hat{\mathbf{y}}|^2$ and $|\hat{\mathbf{y}}|^2$ are given. Phase retrieval with background information can be regarded as one of the special type of utilizing the inference signal $\mathbf{y}$. It only requires the information of $\mathbf{y}$ and $|\hat{\mathbf{x}}+\hat{\mathbf{y}}|^2$. Besides, the uniqueness about this model can exclude all the other trivial and nontrivial solutions. 
\section{Algorithm for the phase retrieval with background information} 
In this section, methods will be proposed to search for the solution of (2). Assuming the conditions in Theorem 2.3 are satisfied besides $m=n+k$. We can certainly solve the linear equations (10) to get the solution. But with the increasing of $n$, the cost for solving the linear equations becomes quite huge. Besides, the construction of $\mathbf{H}$ and $\mathbf{T}$ is complex especially for 2-D signal.\\
\indent
The main idea of our method is to construct an objective function $f(\mathbf{u})$ where the global optimum is the solution of (2), then utilizing the projected gradient descent method to search for the ground truth. The optimization problem in the paper is:
\begin{eqnarray}
	\operatorname*{minimize}\limits_{\mathbf{u}\in\mathbb{R}^{n+k}}~~&f(\mathbf{u})=\frac{1}{2(n+k)}\sum_{i=1}^{n+k}(|\mathbf{F}_i ^*\mathbf{u}|-b_i^{\frac{1}{2}}) ^2&\\\nonumber
	&\mathbf{u}\in\Omega=\{\mathbf{u}\in\mathbb{R}^{n+k}|u_{n+l}=y_l,~l=1,\cdots,k\}.&
\end{eqnarray}  
Note that $f(\mathbf{u})$ is non-smooth. We calculate its generalized gradient $\partial f(\mathbf{u})$ \cite{Clarke1975Generalized} as below:
\begin{eqnarray}
	\partial f(\mathbf{u})=\frac{1}{(n+k)}\sum_{i=1}^{n+k}(|\mathbf{F}_i ^*\mathbf{u}|-b_i)\frac{\partial(|\mathbf{F}_i ^*\mathbf{u}|)}{\partial\mathbf{u}},
\end{eqnarray}
\begin{eqnarray}
	\text{where}~\frac{\partial(|\mathbf{F}_i ^*\mathbf{u}|)}{\partial\mathbf{u}}=\frac{\partial((|\mathbf{F}_i ^*\mathbf{u}|^2)^{\frac{1}{2}})}{\partial\mathbf{u}}=\frac{1}{2|\mathbf{F}_i ^*\mathbf{u}|}(\mathbf{F}_i\mathbf{F}_i^*\mathbf{u}+\overline{\mathbf{F}_i\mathbf{F}_i^*\mathbf{u}}).
\end{eqnarray}
Substituting (15) into (14), we have:
\begin{eqnarray}
	\partial f(\mathbf{u})&=&\frac{1}{2(n+k)}\sum_{i=1}^{n+k}(|\mathbf{F}_i ^*\mathbf{u}|-b_i)\frac{1}{|\mathbf{F}_i ^*\mathbf{u}|}(\mathbf{F}_i\mathbf{F}_i^*\mathbf{u}+\overline{\mathbf{F}_i\mathbf{F}_i^*\mathbf{u}})\\
	&=&\mathbf{u}-\frac{1}{(n+k)}\sum_{i=1}^{n+k}\mathbf{F}_i(b_i\frac{\mathbf{F}_i^*\mathbf{u}}{|\mathbf{F}_i^*\mathbf{u}|})\\
	&=&\mathbf{u}-\tilde{\mathbf{u}},
\end{eqnarray}
where $|\mathbf{F}_i^*\tilde{\mathbf{u}}|=b_i$, $i=1,\cdots,n+k$, besides $\mathbf{F}^*_i\mathbf{u}$ and $\mathbf{F}^*_i\tilde{\mathbf{u}}$ have the same phase. Denote $\frac{\mathbf{F}^*\mathbf{u}}{|\mathbf{F}^*\mathbf{u}|}=\mathbf{1}$ when $|\mathbf{F}^*\mathbf{u}|=\mathbf{0}$.\\
\indent
Then we apply Algorithm 1 to search for the solution of (13). In Algorithm 1, $\lambda$ is the step size calculated by the backtracking method which can be seen in Algorithm 2. \\
\begin{algorithm}[!htb] 
	\renewcommand{\algorithmicrequire}{\textbf{Input:}}
	\renewcommand\algorithmicensure {\textbf{Output:} }
	\caption{ The projected gradient method} 
	\label{alg:Framwork} 
	\begin{algorithmic}[1] 
		\REQUIRE $\{\mathbf{b},\mathbf{y},\varepsilon\}$ ~~\\ 
		$\mathbf{b}$: the Fourier measurement.\\
		$\mathbf{y}$: the background information.\\
		$\varepsilon$:~~the allowed error bound.\\
		\ENSURE ~~\\ 
		$\mathbf{x}^*$: an estimation for the real signal $\mathbf{x}$.\\
		\end{algorithmic}
		\vskip 4mm
		\hrule
		\vskip 2mm
	\noindent\textbf{Initialization:}
	\begin{algorithmic}
		\STATE $\mathbf{u}_0^{'}=\mathcal{F}^{-1}(\mathbf{b}^{\frac{1}{2}})$, where $\mathcal{F}^{-1}$ is the inverse Fourier transform\\
		\STATE $\mathbf{u}_0=\mathbb{P}_{\Omega}(\mathbf{u}_0^{'})$, where $\Omega=\{\mathbf{u}|u_{n+i}=y_i,~i=1,\cdots,k\}$, $\mathbb{P}_{\Omega}$ is the projection operator.\\
	\end{algorithmic}
	\textbf{General Step($p=1,2,\cdots$):}
	\begin{algorithmic}[1] 
		\STATE $\mathbf{u}_{p+1}^{'}=\mathbf{u}_p-\lambda\partial f(\mathbf{u}_p)$, $\lambda$ is determined by the Algorithm 2.\\
		\STATE $\mathbf{u}_{p+1}=\mathbb{P}_{\Omega}(\mathbf{u}_{p+1}^{'})$.\\
		\IF{$\big|\big|\partial f(\mathbf{u}_{p+1})\big|\big|\leq\varepsilon$}
		\STATE $\mathbf{z}^{*}=\mathbf{u}_{p+1}$.\\
		\ENDIF
		\STATE Extracting the first $n$ elements of $\mathbf{z}^*$ to get  $\mathbf{x}^*$ 
	\end{algorithmic}
\end{algorithm}
\begin{algorithm}
	\caption{Stepsize Choosing via Backtracking Method} 
	\label{alg:Framwork} %
	\renewcommand{\algorithmicrequire}{\textbf{Input:}}
	\renewcommand\algorithmicensure {\textbf{Output:} }
	\begin{algorithmic}   
		\REQUIRE$\{\mathit{f}(\mathbf{\mathbf{u}_i}), \partial\mathit{f}(\mathbf{u}_i),\mathbf{u}_{i},\beta\}$ 
		\STATE $\beta\in(0,1)$ is a predetermined parameter
		\ENSURE$\lambda$
		\vskip 4mm
		\hrule
		\vskip 2mm
	\end{algorithmic}
   $\mathbf{General~step}$  
	\begin{algorithmic}[1]   
		\STATE set $\lambda=1$\\
		\STATE Repeat $\lambda\leftarrow 0.5\lambda$ until\\
		$\mathit{f}(\mathbf{u}_{i}-\lambda\partial\mathit{f}(\mathbf{u}_{i}))\leq\mathit{f}(\mathbf{u}_{i})-\lambda\beta||\partial\mathit{f}(\mathbf{u}_{i})||^2$
	\end{algorithmic}  
\end{algorithm} 
\indent
Finding the local optimum in the non-convex problem is NP hard. Now, there is few global convergence guarantee for phase retrieval algorithm based on the Fourier transform such as gradient descent method or error reduction method. As a result, we can't also guarantee the convergence of these algorithms because of the non-convex property\cite{Shechtman2015Phase}. But we can prove that if we find a stationary point in Algorithm 1, it is the global optimum with probability 1 which can largely relief the ill condition of phase retrieval.
\begin{theorem}
	Suppose $m=n+k$, $n\geq 1$, besides 
	\begin{eqnarray*}
		k\geq
		\left\{
		\begin{aligned}
			&3n-2,~m-2n+1\text{~is odd}&\\
			&3n-1,~m-2n+1\text{~is even}&
		\end{aligned}
		\right..
	\end{eqnarray*} $y_i\overset{i.i.d}{\sim}\mathcal{N}(\mu,\sigma^2),~i=1,\cdots,k$. If a stationary point for (13) is found by Algorithm 1, then this stationary point is the global optimum with probability 1. 
\end{theorem}
Proof. If a stationary point $\mathbf{u}_{p}$ is found by Algorithm 1. Then $\partial f(\mathbf{u}_{p})=0$. Combined with (18), we can have, 
\begin{eqnarray}
	\partial f(\mathbf{u}_{p})=\mathbf{u}_p-\tilde{\mathbf{u}}=\mathbf{0}.
\end{eqnarray}
So $\tilde{\mathbf{u}}\in\Omega$. On the other hand, because $|\mathbf{F}_i^*\tilde{\mathbf{u}}|=b_i,i=1,\cdots,n+k$ which satisfy the constraints in the Fourier domain. Then utilizing Theorem 2.4, $\mathbf{u}_{p}=\tilde{\mathbf{u}}$ is the only global optimum for (13) with probability 1.\hfill$\blacksquare$
\section{Numerical Simulations} 
In this section, we make simulations to test our model together with Algorithm 1. In all tests, the background information is generated by Gaussian distribution. All these tests are carried out on the Lenovo desktop with a 3.60 GHz Intel Corel i7 processor and 4GB DDR3 memory. The source code can be found in:\url{https://github.com/Ziyang1992/Phase-retrieval.git}. The relative error is defined as below:
\begin{eqnarray}
	\frac{||\mathbf{x}^*-\mathbf{x}||}{||\mathbf{x}||},
\end{eqnarray} 
where $\mathbf{x}^*$ is the estimation for the real solution $\mathbf{x}$. The length of the 1-D signals used in this test is 100. The size of the 2-D pictures utilized is $512\times 512$. The maximum iteration of the projected gradient descent method for the 1-D signals and 2-D pictures is 300 and 100 respectively.\\ 
\indent We denote $\beta=0.2$ in our simulations. Notice that the initial estimation for the step size is $\lambda=1$ in Algorithm 2, then the stop criterion is satisfied automatically:
\begin{eqnarray}
	f(\mathbf{u}_p-\partial f(\mathbf{u}_p))=f(\tilde{\mathbf{u}}_p)=0\leq0.3||\mathbf{u}_p-\tilde{\mathbf{u}}_p||^2= f(\mathbf{u}_p)-0.2||\mathbf{u}_p-\tilde{\mathbf{u}}_p||^2,  
\end{eqnarray} 
where the last equality is derived by the Parseval equality.\\
\indent
As a result, Algorithm 2 directly runs the step 3, so $\lambda=1$ in all the iterations.
Especially, if $\lambda=1$ in all the iterations, the projected gradient method is actually the same with the error reduction method. There is a similar conclusion in \cite{Fienup1982Phase}.
\subsection{The effects of the length of the background information}
\indent
First, we will test the effect of $k$ on the performance of Algorithm 1. 1-D signals and 2-D pictures are considered respectively. For the 1-D signal, the ratios between $k$ and $n$ are 2:0.1:8 with ':' the matlab notation. At each ratio, we test three different types of 1-D signal besides generating background information and record the combinational Fourier power spectrum $\mathbf{b}$. The tests are replicated 100 times. The types of these three signals are below:
\begin{itemize}
	\item Type 1: $\mathbf{x}\overset{\text{i.i.d}}{\sim}\mathcal{N}(\mathbf{0},\mathbf{I})$
	\item Type 2: the annual mean global surface temperature anomaly which can be download in \url{http://rcada.ncu.edu.tw/research1_clip_ex.htm}. 
	\item Type 3:
	\begin{eqnarray*}
		f(t)=\text{cos}(39.2\pi t-12\text{sin}2\pi t)+\text{cos}(85.4\pi t+12\text{sin}2\pi t),~t\in(0,1)
	\end{eqnarray*}
\end{itemize}
We utilize Algorithm 1 to obtain an estimation. If the relative error is below $10^{-5}$, we judge it a success. The mean recovery rate under each ratio is the overall successful times divided by 100. The result is shown in Figure 2.\\
\begin{figure}
	\centering
	\includegraphics[width=3in]{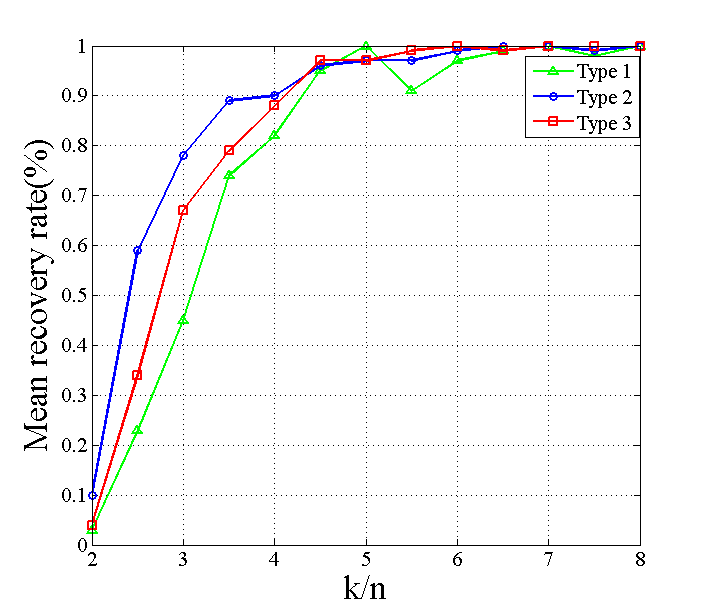}
	\caption{The mean recovery rate for the 1-D signal.}
\end{figure}
\indent
From Figure 2, we can find that the recovery rates of all three types of signal gradually increase when obtaining more background information. When $k/n\geq4$, all the mean recovery rates are greater than 80\%. When $k/n\geq6$, the mean recovery rates are all nearly 100\%. In most of the cases, the mean recovery rates of structure signals(type 2 and type 3) are higher than random signal(type 1). At $k/n=3$ which is around the theoretical limits in this paper, the least mean recovery rate of the signals is about 40\%. It demonstrates that there is a gap between the practice and the theory.\\
\indent
For the 2-D signal,  we utilize two criteria to evaluate the recovery quality. The first one is the Peak Signal to Noise Ratio(PSNR). The second is the Similarity Structural index(SSIM)\cite{1284395} which is between 0 and 1. The SSIM of the high quality recovery picture is closer to 1. The test pictures are Lenna, Baboon, BARB and Harbour. At each dimension of the pictures, the background information is $k/n$ folds than the object. At each $k/n$ ratio, we record the recovery pictures' mean PSNR and mean SSIM under 100 different background information. All the pictures are in the center of the background information. The results are shown in Figure 3.\\
\begin{figure}[!htp]	
	\begin{subfigure}[t]{3in}			
		\includegraphics[width=3in]{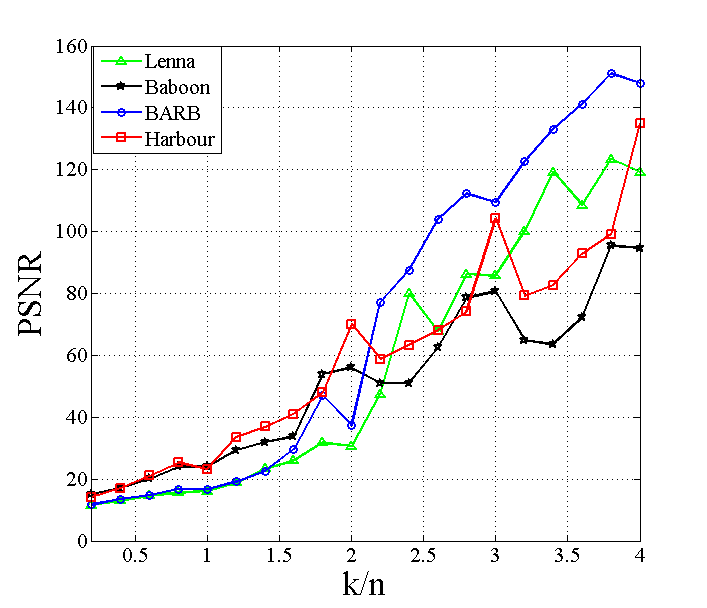}
		\caption{PSNR.}\label{fig:1a}	
	\end{subfigure}
	\begin{subfigure}[t]{3in}			
		\includegraphics[width=3in]{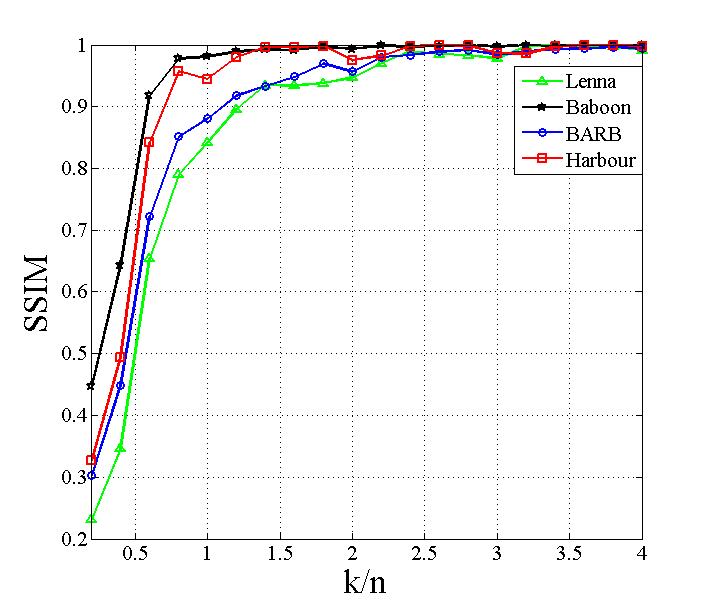}
		\caption{SSIM.}\label{fig:1a}	
	\end{subfigure}	
	\caption{Recovery at each $k/n$.}\label{fig:1}
\end{figure}
\indent
From Figure 3 we can find that the quality of the pictures recovered is also gradually increasing with $k/n$ getting large. When $k/n\geq1.5$ namely the whole background information is 5 folds more than the objects, the least mean SSIM is above 0.9 and the least mean PSNR is above 20. Notice when $k/n=1$, the whole background information is 3 folds than the object which is close to the theoretical limit in Theorem 2.4. At this ratio, the least mean PSNR is nearly 20 and the least mean SSIM of it is more than 0.8 which means the high quality of the recovered pictures. But for the 1-D signals, the least mean recovery rate is about $45$\% when the background information is 3 folds than the original signal. This fully demonstrates the high dimension of the signal can relief the ill condition of the phase retrieval problem.  
\subsection{Influenced by the location or the bias of the background information}
\indent
In this test, the object is 2-D picture Baboon. First, we will test how the location of the background information influences the performance of the algorithm. $k/n=2$. From left corner to the center in Figure 4(a), $17\time 17$ different positions of Baboon are considered. At each position, we replicate the test with 10 times. In Figure 4(b), 4(c) and 4(d), the coordinate of every single pixel means the location of the left top of the Baboon in the whole combined picture. Due to the symmetry of the position, we only consider the four ninths parts in Figure 4(a). We can see that both three indexes(mean PSNR, mean SSIM and mean relative error) of the Baboon recovered in the center are comparatively better from Figure 4. So 2-D objects are in the central of the background information in the following tests.\\
\begin{figure}	
	\begin{subfigure}[t]{1.5in}
		\includegraphics[width=1.1in]{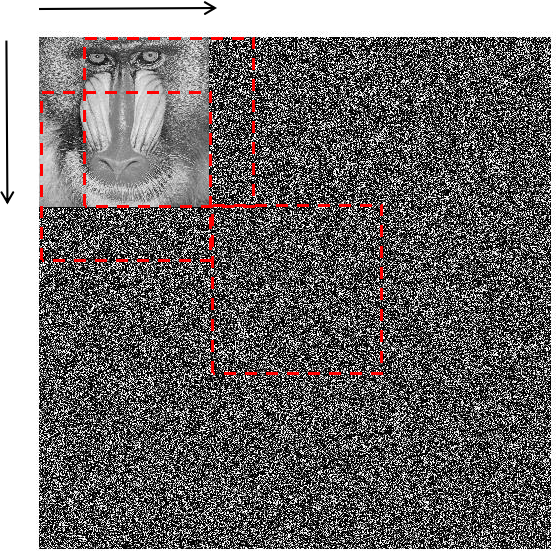}
		\caption{Different positions of Baboon.}\label{fig:1b}
	\end{subfigure}
	\begin{subfigure}[t]{1.5in}
		\includegraphics[width=1.3in]{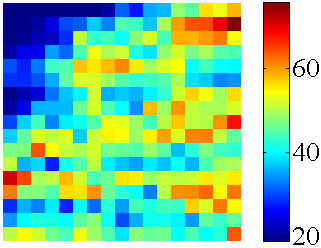}
		\caption{PSNR.}\label{fig:1b}
	\end{subfigure}
	\begin{subfigure}[t]{1.5in}			
		\includegraphics[width=1.4in]{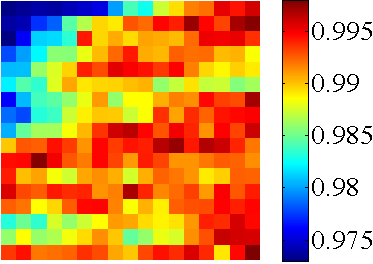}
		\caption{SSIM.}\label{fig:1a}	
	\end{subfigure}
	\begin{subfigure}[t]{1.5in}			
		\includegraphics[width=1.3in]{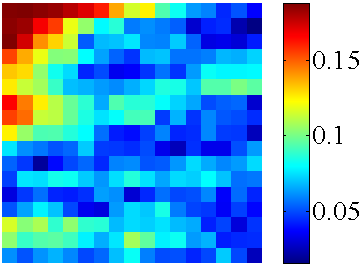}
		\caption{Relative error.}\label{fig:1a}		
	\end{subfigure}
	\caption{Recovery in different positions.}\label{fig:1}
\end{figure}
\indent
In the practice, the background information may have bias. Thus in this simulation test, we assume the real background information have a Gaussian bias with mean zero and variance $\sigma^2$ comparing to the prior background information. The length of the background information is four folds of the length of the image. Under different $\sigma^2$, we make tests with 100 times and record the mean PSNR, SSIM and relative error. The results are shown in Figure 5 and Table 1.\\
\indent
We can find that the algorithm can resist the bias of the background information in some degree. Though the relative error is 0.18 for $\sigma=0.1$, we can also distinguish the image.
\begin{figure}	
	\begin{subfigure}[t]{1.5in}
		\includegraphics[width=1.3in]{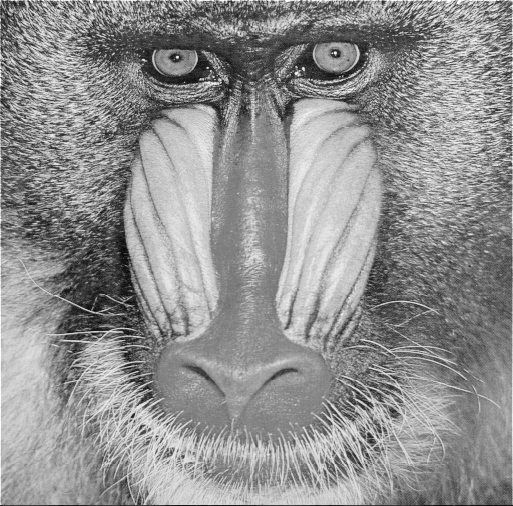}
		\caption{benchmark.}\label{fig:1b}
	\end{subfigure}
	\begin{subfigure}[t]{1.5in}			
		\includegraphics[width=1.3in]{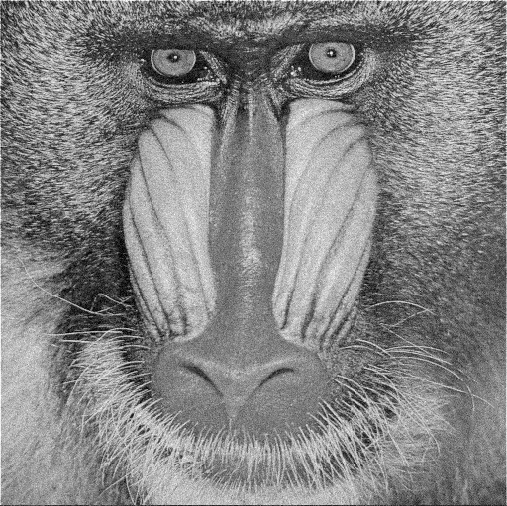}
		\caption{$\sigma=0.05$.}\label{fig:1a}		
	\end{subfigure}
	\begin{subfigure}[t]{1.5in}			
		\includegraphics[width=1.3in]{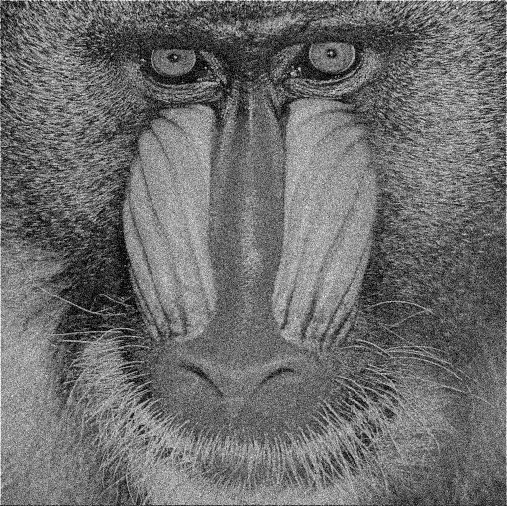}
		\caption{$\sigma=0.1$.}\label{fig:1a}		
	\end{subfigure}
	\begin{subfigure}[t]{1.5in}			
		\includegraphics[width=1.3in]{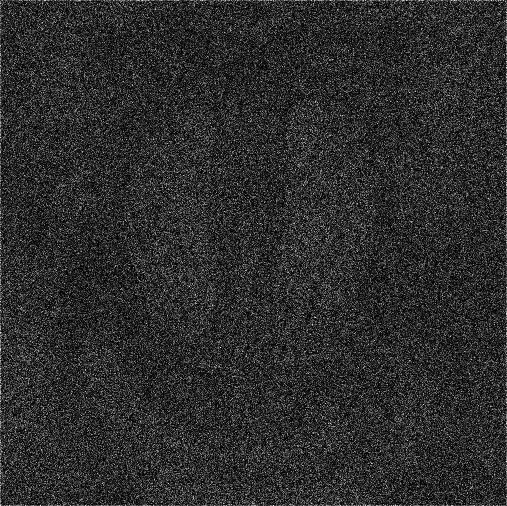}
		\caption{$\sigma=0.5$.}\label{fig:1a}		
	\end{subfigure}
	\caption{Pictures recovered under different bias of the background information.}\label{fig:1}
\end{figure}
\begin{table}
	\centering
	\caption{The recovery criteria under different bias of the background information.}
	\vspace{0.2cm}
	\begin{tabular}{cccc}
		\hline
		\multicolumn{1}{c}{}$\sigma$~~  & PSNR &SSIM&Relative error\\
		\hline
		$0.05$ &25.55&0.77&0.09 \\
		$0.1$ &19.34&0.53&0.18\\
		$0.5$ &0.99&0.02&1.51\\
		\hline
	\end{tabular}
\end{table}
\subsection{Sensitivity to the measurement noise} 
In this test, we will test the algorithm's ability to resist the measurement noise for the 1-D signal and 2-D picture. The noise model is as below:
\begin{eqnarray}
	\sqrt{b_i}=|\mathbf{F}_i^*\mathbf{z}+\varepsilon_i|,~i=1,\cdots,m,
\end{eqnarray}
which is used in \cite{Netrapalli2013Phase}, where $\varepsilon_i$ is the Gaussian noise.\\
\indent
Assume there is no bias for the background information. For the 1-D random signal, the Signal Noise Ratio(SNR) is 0dB:5dB:60dB. At each SNR we will generate one signal with background information and 100 different Gaussian measurement noise. $k/n=7$ besides the maximum iteration of the algorithm is 500. From Figure 6 we can find the projected gradient method can be robust to the measurement noise. When SNR is above $40\text{dB}$, the relative error is below $10^{-2}$. \\
\begin{figure}
	\centering		
	\includegraphics[width=3in]{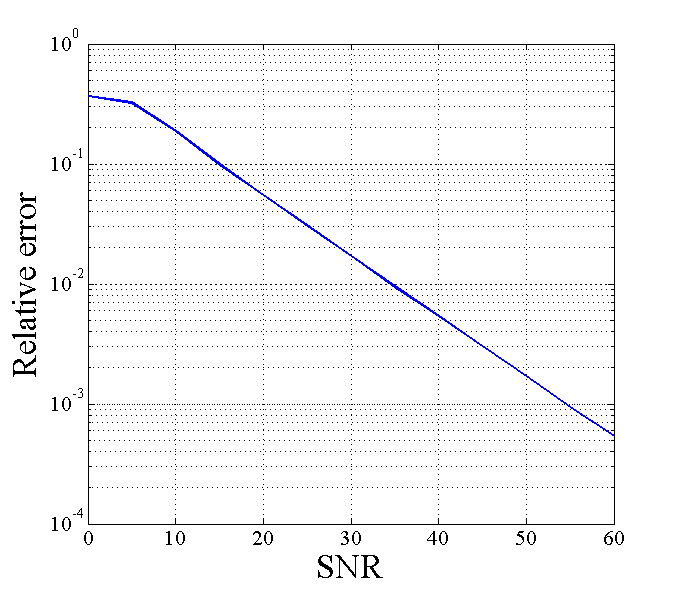}
	\caption{The relative error under each SNR.}\label{fig:1a}
\end{figure}
\indent
For the 2-D pictures, we choose five single channel images as objects. The background information in each dimension is four folds of the image's length. From table 2, we can see that the method can also resist the Gaussian measurement noise for the 2-D pictures. The relative error isn't large besides images recovered can be distinguished well.\\
\indent
All in all, the projected gradient descent method can have a good performance for the phase retrieval with background information no matter to deal with 1-D signals or 2-D pictures.
\begin{table}
	\centering
	\caption{The robustness test of the method for 2-D pictures.}
	\vspace{0.01cm}
	\scalebox{0.73}{
		\begin{tabular}{ccccccccc}
			\hline
			\multicolumn{1}{c}{} &\multicolumn{2}{c}{PSNR}& & \multicolumn{2}{c}{SSIM}& &\multicolumn{2}{c}{Relative error}\\
			\cline{2-3}\cline{5-6}\cline{8-9}
			\multicolumn{1}{c}{}&$\sigma=0.003$ & $\sigma=0.001$ & &$\sigma=0.003$ & $\sigma=0.001$& & $\sigma=0.003$&$\sigma=0.001$ \\
			\hline
			Baboon &11.93&21.54& &0.35&0.74&&0.43&0.14\\
			Barbara &13.93&24.03& &0.30&0.68&&0.41&0.13\\
			Golden hill &13.90&23.79& &0.24&0.67&&0.41&0.13\\
			Harbour&12.81&22.50& &0.26 &0.62& &0.43 &0.14 \\
			Standard Lenna&13.91 &23.26 & &0.19 &0.55& &0.40&0.14\\
			\hline
	\end{tabular}}
\end{table}
\section{Conclusion}
In this paper, we come up with a model called phase retrieval with background information. We prove that with sufficient background information the solution of the phase retrieval problem becomes unique with large probability which largely reliefs the ill condition of phase retrieval. Based on this model, quadratic loss function is constructed and we apply projected gradient descent method to search for the ground truth. We prove the stationary point is the global optimum with probability 1. In numerical tests, the method can have a good performance for 1-D signals and 2-D pictures. At the same time, it can have a good performance although suffering from the bias of the background information and corruption by the Gaussian noise. \\
\indent
There are several works we will be keen to do in the future work. First, the length of the background information for the uniqueness in the Theorem 2.4 is a sufficient condition. In the numerical test, especially for the 2-D signal, the method can also easily converge to the ground truth below this limitation. As a result, we will develop theories to decrease the acquired length of the background information.
Second, the stationary point of the quadratic function with sufficient background information is the global optimum which largely reliefs the difficulty of the non-convex and non-smooth problem. So we will be keen to utilize new optimization methods to escape from these fixed points and converge to the stationary point more efficiently besides making corresponding convergence analyses.
\section{Acknowledgement}
This work is supported in part by National Natural Science foundation(China): 61571008.
\bibliography{1.bib}

\begin{thebibliography}{10}

\bibitem{Miao1999Extending}
Jianwei Miao, Pambos Charalambous, Janos Kirz, and David Sayre.
\newblock Extending the methodology of x-ray crystallography to allow imaging
  of micrometre-sized non-crystalline specimens.
\newblock {\em Nature}, 400(6742):342--344, 1999.

\bibitem{fienup1987phase}
C~Fienup and J~Dainty.
\newblock Phase retrieval and image reconstruction for astronomy.
\newblock {\em Image Recovery: Theory and Application}, pages 231--275, 1987.

\bibitem{Zheng2013Wide}
G.~Zheng, R~Horstmeyer, and C.~Yang.
\newblock Wide-field, high-resolution fourier ptychographic microscopy.
\newblock {\em Nature Photonics}, 7(9):739--745, 2013.

\bibitem{Hofstetter1964Construction}
E.~M Hofstetter.
\newblock Construction of time-limited functions with specified autocorrelation
  functions.
\newblock {\em IEEE Transactions on Information Theory}, 10(2):119--126, 1964.

\bibitem{Hayes1982The}
M.~H Hayes.
\newblock The reconstruction of a multidimensional sequence from the phase or
  magnitude of its fourier transform.
\newblock {\em IEEE Transactions on Acoustics Speech and Signal Processing},
  30(2):140--154, 1982.

\bibitem{Gerchberg1971A}
R.~W. Gerchberg.
\newblock A practical algorithm for the determination of phase from image and
  diffraction plane pictures.
\newblock {\em Optik}, 35:237--250, 1971.

\bibitem{Fienup1982Phase}
J~R Fienup.
\newblock Phase retrieval algorithms: a comparison.
\newblock {\em Applied Optics}, 21(15):2758--2769, 1982.

\bibitem{Beinert2015Ambiguities}
Robert Beinert and Gerlind Plonka.
\newblock Ambiguities in one-dimensional discrete phase retrieval from fourier
  magnitudes.
\newblock {\em Journal of Fourier Analysis and Applications}, 21(6):1169--1198,
  2015.

\bibitem{6853917}
H.~Ohlsson and Y.~C. Eldar.
\newblock On conditions for uniqueness in sparse phase retrieval.
\newblock In {\em 2014 IEEE International Conference on Acoustics, Speech and
  Signal Processing (ICASSP)}, pages 1841--1845, 2014.

\bibitem{JURi2013Unique}
Juri Ranieri, Amina Chebira, Yue~M. Lu, and Martin Vetterli.
\newblock Phase retrieval for sparse signals: Uniqueness conditions.
\newblock {\em ArXiv Preprint arXiv:1308.3058}, 2013.

\bibitem{Shechtman2013GESPAR}
Yoav Shechtman, Andre Beck, and Yonina~C. Eldar.
\newblock Gespar: Efficient phase retrieval of sparse signals.
\newblock {\em IEEE Transactions on Signal Processing}, 62(4):928--938, 2013.

\bibitem{Eldar2014Sparse}
Yonina~C. Eldar, Pavel Sidorenko, Dustin~G. Mixon, Shaby Barel, and Oren Cohen.
\newblock Sparse phase retrieval from short-time fourier measurements.
\newblock {\em IEEE Signal Processing Letters}, 22(5):638--642, 2014.

\bibitem{Bendory2017Phase}
Tamir Bendory and Yonina~C. Eldar.
\newblock Phase retrieval from stft measurements via non-convex optimization.
\newblock In {\em IEEE International Conference on Acoustics, Speech and Signal
  Processing}, pages 4770--4774, 2017.

\bibitem{Kim1990Iterative}
W~Kim and M.~H Hayes.
\newblock Iterative phase retrieval using two fourier transform intensities.
\newblock In {\em International Conference on Acoustics, Speech, and Signal
  Processing}, pages 1563--1566, 1990.

\bibitem{Kim2002Phase}
W.~Kim and M.~H. Hayes.
\newblock Phase retrieval using a window function.
\newblock {\em IEEE Transactions on Signal Processing}, 41(3):1409--1412, 2002.

\bibitem{Raz2013Vectorial}
Oren Raz, Nirit Dudovich, and Boaz Nadler.
\newblock Vectorial phase retrieval of 1-d signals.
\newblock {\em IEEE Transactions on Signal Processing}, 61(7):1632--1643, 2013.

\bibitem{Beinert2017Enforcing}
Robert Beinert and Gerlind Plonka.
\newblock Enforcing uniqueness in one-dimensional phase retrieval by additional
  signal information in time domain.
\newblock {\em ArXiv Preprint arXiv:1604.04493}, 2016.

\bibitem{Beinert2015}
Robert Beinert and Gerlind Plonka.
\newblock Ambiguities in one-dimensional discrete phase retrieval from fourier
  magnitudes.
\newblock {\em Journal of Fourier Analysis and Applications}, 21(6):1169--1198,
  Dec 2015.

\bibitem{Nawab1983Signal}
Hamid Nawab, T.~F Quatieri, and J.~S Lim.
\newblock Signal reconstruction from the short-time fourier transform
  magnitude.
\newblock {\em IEEE Transactions on Acoustics Speech and Signal Processing},
  31(4):986--998, 1983.

\bibitem{Pan2012Condition}
Victor~Y. Pan and Guoliang Qian.
\newblock Condition numbers of random toeplitz and circulant matrices.
\newblock {\em ArXiv Preprint arXiv:1212.4551}, 2012.

\bibitem{Clarke1975Generalized}
Frank~H Clarke.
\newblock Generalized gradients and applications.
\newblock {\em Transactions of the American Mathematical Society},
  205(205):247--262, 1975.

\bibitem{Shechtman2015Phase}
Yoav Shechtman, Yonina~C. Eldar, Oren Cohen, and Henry~Nicholas Chapman.
\newblock Phase retrieval with application to optical imaging: A contemporary
  overview.
\newblock {\em IEEE Signal Processing Magazine}, 32(3):87--109, 2015.

\bibitem{1284395}
Zhou Wang, A.~C. Bovik, H.~R. Sheikh, and E.~P. Simoncelli.
\newblock Image quality assessment: from error visibility to structural
  similarity.
\newblock {\em IEEE Transactions on Image Processing}, 13(4):600--612, 2004.

\bibitem{Netrapalli2013Phase}
Praneeth Netrapalli, Prateek Jain, and Sujay Sanghavi.
\newblock Phase retrieval using alternating minimization.
\newblock {\em IEEE Transactions on Signal Processing}, 63(18):4814--4826,
  2013.

\end{thebibliography}



\end{document}